\RequirePackage{fix-cm}

\documentclass{bmcart}

\usepackage[utf8]{inputenc} %unicode support
\usepackage{amssymb}
\usepackage{graphicx}
\usepackage{natbib}
\usepackage{amsmath}
\usepackage[footnotesize,sl,SL,hang,tight]{subfigure}  % helpful package for aligning figures next to each other
\usepackage{color}   % {\color{red} TEXT }
\usepackage{comment}
\usepackage{enumitem}
\usepackage{url}
\newcommand{\todo}[1]{}

\startlocaldefs
\endlocaldefs

\begin{document}
%\setlistdepth{10}

%%% Start of article front matter
\begin{frontmatter}

\begin{fmbox}
\dochead{Research}


\title{Exploiting citation networks for large-scale author name disambiguation}

%%%%%%%%%%%%%%%%%%%%%%%%%%%%%%%%%%%%%%%%%%%%%%
%%                                          %%
%% Enter the authors here                   %%
%%                                          %%
%% Specify information, if available,       %%
%% in the form:                             %%
%%   <key>={<id1>,<id2>}                    %%
%%   <key>=                                 %%
%% Comment or delete the keys which are     %%
%% not used. Repeat \author command as much %%
%% as required.                             %%
%%                                          %%
%%%%%%%%%%%%%%%%%%%%%%%%%%%%%%%%%%%%%%%%%%%%%%

\author[
   addressref={aff1},                   % id's of addresses, e.g. {aff1,aff2}
%   corref={aff1},                       % id of corresponding address, if any
%   noteref={n1},                        % id's of article notes, if any
%   email={jane.e.doe@cambridge.co.uk}   % email address
]{\inits{C}\fnm{Christian} \snm{Schulz}}
\author[
   addressref={aff1},
%   email={jane.e.doe@cambridge.co.uk}   % email address
]{\inits{A}\fnm{Amin} \snm{Mazloumian}}
\author[
   addressref={aff2},
%   email={jane.e.doe@cambridge.co.uk}   % email address
]{\inits{AM}\fnm{Alexander M} \snm{Petersen}}
\author[
   addressref={aff2},
%   email={jane.e.doe@cambridge.co.uk}   % email address
]{\inits{O}\fnm{Orion} \snm{Penner}}
\author[
   addressref={aff1},
   corref={aff1},                       % id of corresponding address, if any
   email={dhelbing@ethz.ch}   % email address
]{\inits{D}\fnm{Dirk} \snm{Helbing}}

%%%%%%%%%%%%%%%%%%%%%%%%%%%%%%%%%%%%%%%%%%%%%%
%%                                          %%
%% Enter the authors' addresses here        %%
%%                                          %%
%% Repeat \address commands as much as      %%
%% required.                                %%
%%                                          %%
%%%%%%%%%%%%%%%%%%%%%%%%%%%%%%%%%%%%%%%%%%%%%%

\address[id=aff1]{%                           % unique id
  \orgname{ETH Zurich, Department of Humanities and Social Sciences, Chair of Sociology, in particular of Modeling and Simulation}, % university, etc
  \street{Clausiusstrasse 50},                     %
  \postcode{CH-8092}                                % post or zip code
  \city{Zurich},                              % city
  \cny{Switzerland}                                    % country
}
\address[id=aff2]{%
  \orgname{IMT Institute for Advanced Studies Lucca},
  \street{Piazza San Francesco 19},
  \postcode{55100}
  \city{Lucca},
  \cny{Italy}
}

%%%%%%%%%%%%%%%%%%%%%%%%%%%%%%%%%%%%%%%%%%%%%%
%%                                          %%
%% Enter short notes here                   %%
%%                                          %%
%% Short notes will be after addresses      %%
%% on first page.                           %%
%%                                          %%
%%%%%%%%%%%%%%%%%%%%%%%%%%%%%%%%%%%%%%%%%%%%%%

\begin{artnotes}
%\note{Sample of title note}     % note to the article
%\note[id=n1]{Equal contributor} % note, connected to author
\end{artnotes}

\end{fmbox}% comment this for two column layout

%%%%%%%%%%%%%%%%%%%%%%%%%%%%%%%%%%%%%%%%%%%%%%
%%                                          %%
%% The Abstract begins here                 %%
%%                                          %%
%% Please refer to the Instructions for     %%
%% authors on http://www.biomedcentral.com  %%
%% and include the section headings         %%
%% accordingly for your article type.       %%
%%                                          %%
%%%%%%%%%%%%%%%%%%%%%%%%%%%%%%%%%%%%%%%%%%%%%%

\begin{abstractbox}
\begin{abstract}
We present a novel algorithm and validation method for disambiguating author names in very large bibliographic data sets and apply it to the full Web of Science (WoS) citation index. Our algorithm relies only upon the author and citation graphs available for the whole period covered by the WoS. A pair-wise publication similarity metric, which is based on common co-authors, self-citations, shared references and citations, is established to perform a two-step agglomerative clustering that first connects individual papers and then merges similar clusters. This parameterized model is optimized using an $h$-index based recall measure, favoring the correct assignment of well-cited publications, and a name-initials-based precision using WoS metadata and cross-referenced Google Scholar profiles. Despite the use of limited metadata, we reach a recall of 87\% and a precision of 88\% with a preference for researchers with high $h$-index values. 47 million articles of WoS can be disambiguated on a single machine in less than a day. We develop an h-index distribution model, confirming that the prediction  is in excellent agreement with the empirical data, and  yielding insight into the utility of the h-index in real academic ranking scenarios.
\end{abstract}

%%%%%%%%%%%%%%%%%%%%%%%%%%%%%%%%%%%%%%%%%%%%%%
%%                                          %%
%% The keywords begin here                  %%
%%                                          %%
%% Put each keyword in separate \kwd{}.     %%
%%                                          %%
%%%%%%%%%%%%%%%%%%%%%%%%%%%%%%%%%%%%%%%%%%%%%%


% MSC classifications codes, if any
%\begin{keyword}[class=AMS]
%\kwd[Primary ]{}
%\kwd{}
%\kwd[; secondary ]{}
%\end{keyword}

\end{abstractbox}
%
%\end{fmbox}% uncomment this for twcolumn layout

\end{frontmatter}

%%%%%%%%%%%%%%%%%%%%%%%%%%%%%%%%%%%%%%%%%%%%%%
%%                                          %%
%% The Main Body begins here                %%
%%                                          %%
%% Please refer to the instructions for     %%
%% authors on:                              %%
%% http://www.biomedcentral.com/info/authors%%
%% and include the section headings         %%
%% accordingly for your article type.       %%
%%                                          %%
%% See the Results and Discussion section   %%
%% for details on how to create sub-sections%%
%%                                          %%
%% use \cite{...} to cite references        %%
%%  \cite{koon} and                         %%
%%  \cite{oreg,khar,zvai,xjon,schn,pond}    %%
%%  \nocite{smith,marg,hunn,advi,koha,mouse}%%
%%                                          %%
%%%%%%%%%%%%%%%%%%%%%%%%%%%%%%%%%%%%%%%%%%%%%%

%%%%%%%%%%%%%%%%%%%%%%%%% start of article main body

\section{Introduction}
\label{intro}

The ambiguity of author names is a major barrier to the analysis of large scientific publication databases on the level of individual researchers~\cite{smalheiser2009author, ferreira2012brief}. Within such databases researchers generally appear only as they appear on any given publication {\it i.e.} by their surname and first name initials. Frequently, however, hundreds or even thousands of individual researchers happen to share the same surname and first name initials. Author name disambiguation is therefore an important prerequisite for the author level analyses of publication data. While many important and interesting problems can be examined without individual level data~\cite{mazloumian2013global, radicchi2012science} a great many other require such data to get to the real heart of the matter. Good examples include the role of gender in academic career success~\cite{sugimoto2013gender}, whether ideas diffuse through the popularity of individual publications or the reputation of the authors~\cite{mazloumian2011citation, petersen2013reputation}, how the specific competencies and experience of the individual authors recombine to search the space of potential innovations~\cite{fleming2004science, fleming2007collaborative}, and whether one can predict scientific carriers~\cite{acuna2012predicting, mazloumian2012predicting, penner2013commentary, penner2013case, penner2013PCP}. Indeed, the importance of getting individual level data has been widely acknowledged, as can be seen in recent large scale initiatives to create disambiguated researcher databases~\cite{ORCID, VIVO}

Algorithmic author name disambiguation is challenging for two reasons. First, existing disambiguation algorithms have to rely on metadata beyond name to distinguish between authors with the same name, much like some administrative institutions do when they distinguish citizens with the same name based on attributes such as date and place of birth. However, in existing large-scale publication databases --- such as Thomson Reuters' Web of Science (WoS) --- metadata is often sparse, especially for older publications. 
Second, disambiguation algorithms may draw false conclusions when faced with incomplete metadata. For instance, when researchers change disciplines they transition to an entirely different part of the citation graph. Therefore, disambiguation algorithms that heavily rely on journal metadata to reconstruct researchers' career trajectories can easily represent such researchers with two different researcher profiles. This issue can be present in any case where an individual metadata (disciplinary profile, collaborators, affiliation) is not consistent over time.

Existing disambiguation algorithms typically exploit metadata like first and middle names, co-authors, publication titles, topic keywords, journal names, and affiliations or email addresses (for an overview see \cite{ferreira2012brief}). Reference~\cite{torvik2005probabilistic} (and enhanced in  \cite{torvik2009author}) presents a comprehensive method that includes all metadata of the MEDLINE database. The use of citation graph data is less common however, since only a few databases include this information. Previous examples to exploit such data include \cite{levin2012citation} which mainly relies on self-citations, and \cite{Walsh2010} that used shared references, but only for the disambiguation of two author names. Both retrieve data from the WoS, which is also used in \cite{d2011heuristic} and \cite{reijnhoudt2013seed+}, however, without exploiting the citation graph. Reference~\cite{d2011heuristic} had access to a manually maintained database of Italian researchers as a gold standard, while \cite{reijnhoudt2013seed+} found a ground truth in Dutch full professor publication lists.

Here, we develop and apply a novel author disambiguation algorithm with the explicit goal of measuring the $h$-index of researchers using the entire WoS citation index database.
Introduced by Hirsch in 2005, the $h$-index is the most widely used measure of an individual's scientific impact. An individual's $h$-index is equal to
the number $h$ of publications that are cited at least $h$ times. It is increasingly used in both informal and formal evaluation and career advancement programs~\cite{ANVUR}.
However, despite its rapidly increasing popularity and use, very little is known about the overall distribution of $h$-indices in science. While an $h$-index of 30 is certainly less frequent than an $h$-index of 20, it is unknown how much less frequent. Models have been developed to estimate the distribution based upon some simple assumptions, but at best, they relied on incomplete data. Perhaps the most straightforward starting point for considering the distribution of $h$-index would be Lotka's law scientific for productivity~\cite{Lotka1926fds}, however in the results section we will show that the empirical data deviates significantly from a Pareto power-law distribution.

The most complete data-centric work to date is that of \cite{Castellano2013}, who calculated a probability distribution $P(h)$ of  $h$-indices using over 30,000 career profiles acquired via Google Scholar. Indeed this work represents a critical step forward in terms of understanding the overall distribution of $h$-indices and the high level dynamics that shape it. However, Google Scholar profiles are biased towards currently active and highly active researchers. As a consequence, their approach may underestimate the number of individuals with  low $h$-index. A proper understanding of the entire $h$-index distribution $P(h)$ is critical to shaping policies and best practices of using it for scientific performance. Furthermore, as research becomes more interdisciplinary, the variation of $h$-index distribution across disciplines must be better understood to prevent biased evaluations. To tackle these and similar challenges, we present an algorithm that is optimized towards reproducing the correct $h$-index of researchers, makes use of the citation network, and is applicable for the entire dataset of WoS.

This manuscript will be laid out in the following manner. First, we will describe our algorithm, novel validation \& optimization approach, and implementation details. Then we will present the results of our optimization procedure and the empirical $h$-index distribution produced by our algorithm. We will compare the empirical distribution to the predictions of a simple theoretical h-index model, which together show excellent agreement.

\section{Methodology}
\label{meth}

\subsection{The Disambiguation Algorithm}
\begin{figure}
\centering{\includegraphics[width=0.95\textwidth]{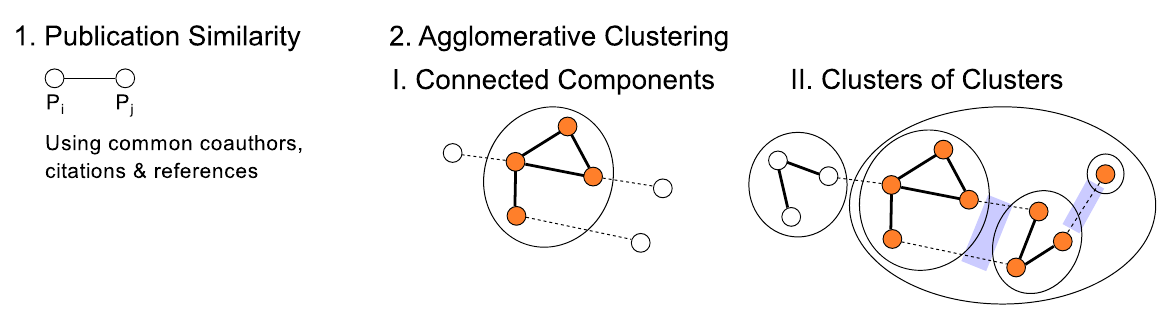}}
\caption{\textbf{For a given set of publications, a measure for publication similarity is used to identify clusters that ideally represent unique researchers.} First, the clustering creates strongly connected components. Second, well-linked clusters are merged.}
\label{fig:schemaAlgorithm}
\end{figure}

As discussed above, the goal of a disambiguation algorithm is to generate sets of publications that can be attributed to specific, individual, researchers. Our algorithm accomplishes this by a two step agglomerative approach (see Fig.~\ref{fig:schemaAlgorithm}).

In the first step the goal is to determine if two papers were likely coauthored by the same individual. To that aim, we are using a similarity score approach to cluster papers. We first calculate the pairwise similarity between all pairs of papers in the dataset of ambiguous names. The similarity score ($s_{ij}$) between two papers $i$ and $j$ is calculated as follows:

\begin{eqnarray}
s_{ij} &=&  \alpha_A \left( \frac{|A_i  \cap  A_j| }{\min \left(\left|A_i\right|,\left|A_j\right|\right)}\right) +  \alpha_S \left( |p_i \cap R_j| + |p_j \cap R_i| \right) + \nonumber \\
& & \alpha_R \left( |R_i  \cap  R_j| \right) + 
\alpha_C \left( \frac{|C_i  \cap  C_j| }{\min \left(\left|C_i\right|,\left|C_j\right|\right)}\right) \label{eq:similarity_fcn}.
\end{eqnarray}
For each paper $p_i$ we denote the reference list as $R_i$; the co-author list as $A_i$; the set of citing papers as $C_i$. Hence in this instantiation of the algorithm, these are the only three pieces of information one must have available for each paper. The $\cap$-operator together with the enclosing  $|$ $|$-operator count the number of common attributes. The first term in Eq.~(\ref{eq:similarity_fcn}) measures the number of co-authors shared by two papers. The second term detects potential self-citations, a well recognized indicator of an increased probability of authorship by the same individual \citep{Hellsten2007}. The third term is the count of common references between the two papers. The fourth term represents the number of papers that cite both publications. The first and last terms are normalized by a technique known as overlap coefficient \citep{salton1968automatic}. It accounts for the higher likelihood of finding similarities when both co-author lists are very long or both publications are well-cited.

Once all pairwise similarities have been calculated, our algorithm moves on to the first of two clustering processes (see Fig.~\ref{fig:schemaAlgorithm}). In this first clustering we start by establishing a link between each pair of papers $(i,j)$, for which the similarity score $s_{ij}$ is greater than a threshold $\beta_1$. Then, each connected component (set of papers that can be reached from each other paper by traversing the previously created links) is labeled as a cluster. The goal is, of course, that all papers in any given cluster belong to one specific author.

In the second clustering process a new similarity score is calculated between all clusters generated in the previous step. Labeling one cluster by $\gamma$ and another by $\kappa$ the similarity between the clusters is calculated as follows:

\begin{equation}
S_{\gamma, \kappa} = \sum_{i \in \gamma j \in \kappa} \frac{s_{ij}\Theta\left( s_{ij} > \beta_2 \right)}{|\gamma||\kappa|}
\label{eq:cluster_merge}.
\end{equation}

Here $|\gamma|$ is the number of publications in cluster $\gamma$, similarly for $|\kappa|$. 
For this step we calculate the similarity between publications in separate clusters.
The overall cluster-cluster similarity is the sum of the $s_{ij}$ similarity weights  that are above a certain threshold $\beta_2$, normalized by the number of papers of the two clusters. A link is then established between the two clusters if the new cluster similarity score $(S_{\gamma, \kappa})$ is greater than a threshold $\beta_3$. Each connected component (set of clusters that can be reached from each other cluster by traversing links) is then merged into a single cluster. Remaining individual papers are added to a cluster if they have a similarity score $s_{ij}$ above a threshold $\beta_4$ with any paper in that cluster. We denote the set of clusters $\{ K_i \}$ finally resulting from our algorithm. Each cluster is a set of papers and should ideally contain all papers published by one specific researcher.

\subsection{Optimization and Validation}

\begin{figure}
	\centering
	\subfigure[]{\label{fig:schemaValidation}\includegraphics[width=0.6038\textwidth]{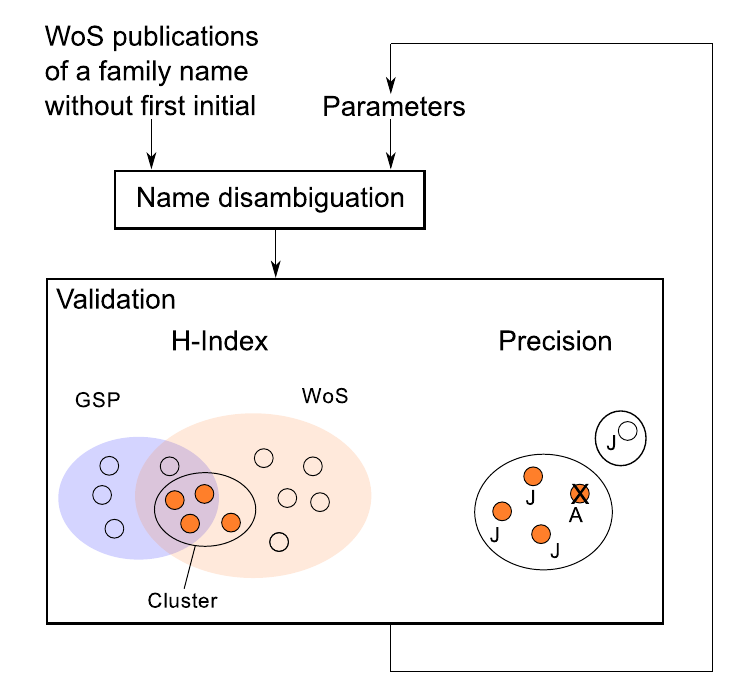}}\hfill
	\subfigure[]{\label{fig:schemaWos}\includegraphics[width=0.3462\textwidth]{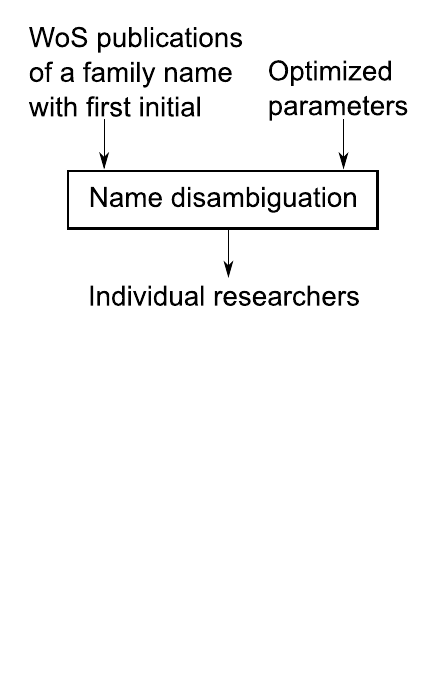}}
	\caption{ \textbf{Optimization and validation procedure.} \subref{fig:schemaValidation} Parameters of the name disambiguation algorithm (shown in Fig. \ref{fig:schemaAlgorithm}) are optimized using Google Scholar Profiles (GSP) for measuring recall and first name initials for measuring precision. \subref{fig:schemaWos} For disambiguating the whole Web of Science (WoS), family names complemented by first initials.}
	\label{fig:schemaValidationAndWos}
\end{figure}

The output of such an algorithm must be validated thoroughly by establishing error rates, specifying their dependence on the size of the researcher profiles produced. Here we develop two techniques for estimating the rates of the two types of statistical errors:  (i) Type I  errors  (``splitting"), which split an individual's publications across two or more clusters, and (ii) Type II errors (``lumping"), which fail to distinguish between two or more author publication sets, i.e. an author mistakenly gets assigned papers from another author. Parameter optimization is a key step in arriving at a functional algorithm (see Fig.~\ref{fig:schemaValidationAndWos}). Our optimization approach differs from many other algorithms in that our optimization procedure does not only seek to minimize ``lumping'' and ``splitting'', but also to optimize an additional specified dimension defined by the research question one wishes to investigate with the disambiguated data. For this work, the dimension of interest is reproducing the $h$-index of individual researchers with high accuracy. Below we describe the details of our algorithm, and then we explain the optimization and validation procedures that we have developed with a specific focus on how to reach the $h$-index accuracy objective.

To assess lumping errors we start by extracting from the WoS database all papers in which a given surname appears in the author field. We then apply our algorithm to this set, ignoring the initials or first names associated with each instance of the given surname. This differs from the typical starting point of previous disambiguation efforts, where the underlying algorithms would be applied to the set of papers in which a given surname together with specific first initial. However, by omitting the first initial information we determine an upper bound for the lumping error, as measured by precision. We define precision of a cluster $i$ which contains various first name initials indexed by $j$:
\begin{equation}
P_i = \frac{\max_{j}\Big(Frequency[FirstNameInitial(j,K_i)]\Big)}{|K_i|} \label{eq:precision}.
\end{equation}
Take the surname ``Smith'', for example. Applying the algorithm to all papers with that surname we get a set of clusters. We can assume that in each cluster the initial that appears on most papers is the ``correct'' initial, and all other initials are likely errors. For example in the cluster where ``J'' is the most frequent initial for ``Smith'' the precision can be estimated as the number of papers with the initial ``J'' divided by the overall number of papers in the cluster. Not all papers with ``J'' may correspond to the same person (``Jason'' versus ``John''), but in the absence of an absolute gold standard this serves as a proxy.

To assess the rate of splitting errors we draw upon Google Scholar Profile (GSP) data. Within an individual's Google Scholar Profile all of an author's publications (indexed by Google Scholar) can be found and we use these profiles as a gold standard. Currently, we have acquired GS profiles for 3,000 surnames. As one would expect, some errors exists within these profiles and papers can be mis-assigned. However, as we discuss below by optimizing for the reconstruction of the $h$-index, this is not a big concern. Before a GSP can be used as a gold standard the contents of the profile must first be cross-referenced to the WoS database by measuring distances in year, title, author list and journal information. A publication is cross-referenced if there is sufficient similarity in multiple fields and if there is no other publication that would also qualify as a match. Once a gold standard publication list has been arrived at, it is straightforward to use it to calculate our algorithm's recall for that profile:
\begin{equation}
R_{\alpha} = \frac{\max\left( |K_i \cap \mathrm{GSProfile}_{\alpha}| \right)}{|\mathrm{GSProfile}_{\alpha} \cap \mathrm{WoS}|} \label{eq:GSP_recall}.
\end{equation}
This is the recall value for a specific GSP (researcher $\alpha$). It corresponds to the percentage of papers in the given profile (that we managed to cross-reference to WoS) that are also in the algorithm-generated cluster which contains most papers of that profile.

The recall value is a measure of how completely we have captured an individual's publication list. However, this does not, necessarily, indicate how well we have captured the portion of an individual's publication list that is relevant to our objective of accurately reproducing the $h$-index. Specifically, when the goal is to measure the $h$-index it is more important to assign every paper that contributes to an individual's $h$-index (the most cited) to his or her cluster, rather than to assign every single paper correctly. Of course, this amplifies the importance of correctly assigning highly cited papers. To measure the extent to which our algorithm can reproduce the $h$-index, we introduce a measure of the $h$-index recall:
\begin{equation}
R^{h}_{\alpha} = \frac{h(\max\left( |K_i \cap \mathrm{GSProfile}_{\alpha}| \right))}{h(|\mathrm{GSProfile}_{\alpha} \cap \mathrm{WoS}|)} \label{eq:h_recall}.
\end{equation}
With the objective of producing the highest quality $h$-index estimates, this measure seamlessly replaces the typical recall measure as a way to evaluate the completeness of clusters. Thus we use it for our optimization and validation procedure instead of Eq.~(\ref{eq:GSP_recall}). However, it is necessary we make clear that in using this $h$-index centric measure the resulting disambiguation is optimized with regards to reproducing $h$-index distribution, but may not be optimal with regards to other criteria. Indeed if a reader were to apply our algorithm, or one like it, with a different goal in mind we advise them to adapt the recall measure to their specific goal.

\subsection{Implementation}

With about 47 million papers (for the analyzed period from 1900 to 2011), 141 million co-author entries, and 526 million citations referring to other articles within the database, the WoS is one of the largest available metadata collections of scientific articles and thus needs to be processed efficiently. While we concentrated on a few features (co-authors and citation graph), our framework can be extended to further metadata as well. We also do not make use of the full citation and co-author network when evaluating a single paper, in the sense that we do not traverse the graph to another paper node which is not directly connected to the paper in question. As a pre-processing step, we compute all publication similarity terms without applying concrete disambiguation parameters. For the complete WoS, we created 4.75 billion links between pairs of papers that have significant similarity and a common name (surname plus first initial). Publication similarity has a computational complexity of $O(n^2)$, where $n$ is the number of papers of the ambiguous name. To reduce the cost of a single paper pair comparison, all information related to a single name is loaded into memory, whereas all feature data (mainly integer IDs) are stored in sorted arrays. For papers that have a publication year difference greater than 5, the computation is skipped to decrease the number of comparisons. This process took 11 hours on standard laptop hardware. Disambiguating the 5.6 million author names, i.e. weighting the similarity links and performing the two-step clustering took less than an hour. For the validation, we kept data for the 500 name networks in memory (consuming less than 4 GB) to test multiple parameter configurations subsequently, so that each parameter test (disambiguation and validation of the 500 names) could be executed in about 5 seconds.

\section{Results}
\label{res}

\begin{figure}
	\centering
	\subfigure[]{\label{fig:random}\includegraphics[width=0.513\textwidth]{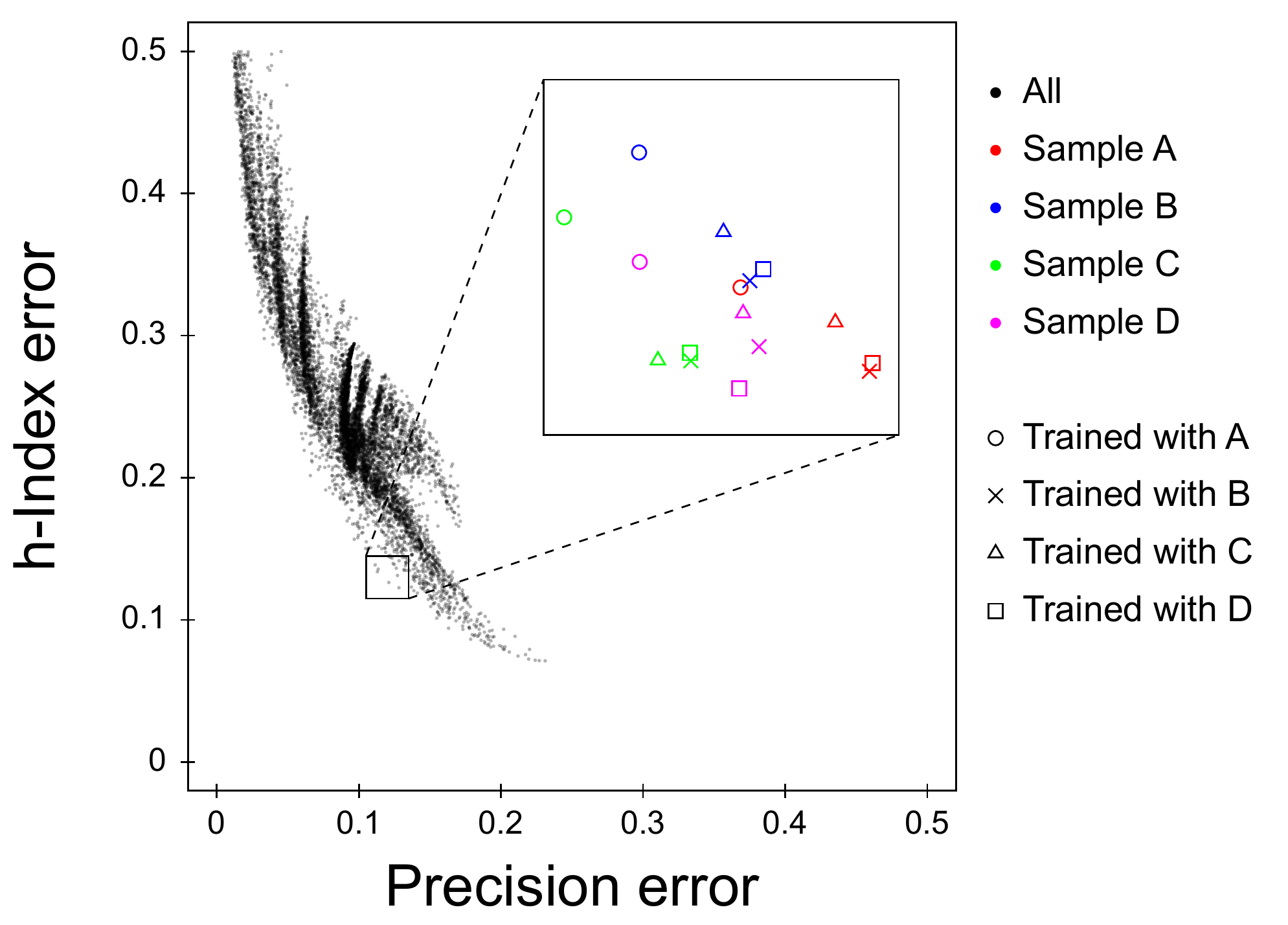}}\hfill
	\subfigure[]{\label{fig:subsets}\includegraphics[width=0.3895\textwidth]{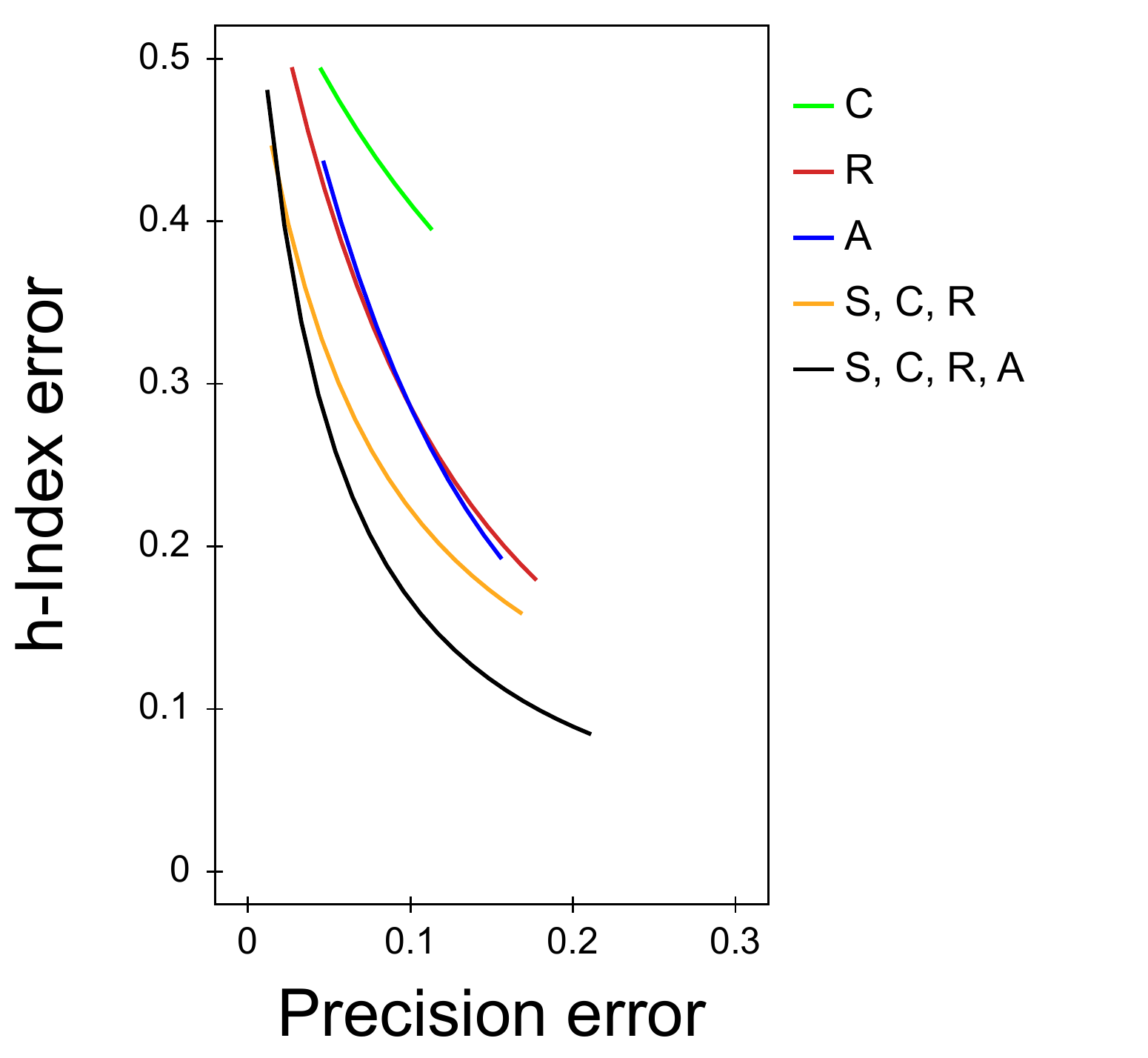}}
	\caption{ \textbf{Optimizing disambiguation parameters.}  \subref{fig:random} 10,000 random disambiguation parameters were tested for the 3,000 family names which we can validate with Google Scholar profiles. Results (indicated as black dots) close to the origin (0,0) yield the best trade-off between precision and h-index correctness. For samples A, B, C and D (consisting of 500 family names each), parameters were further optimized independently and cross-validated. \subref{fig:subsets} Curves represent a lower hull estimate for the results of a random parameter sampling when using only certain features of the metadata (C...Citations, R...References, A...Authors, S...Self-citations). The closer the curves come to the origin, the smaller the error. The combination of all four features lead to the best $h$-index reconstruction.}
	\label{fig:optimization}
\end{figure}

\subsection{Optimizing Disambiguation Parameters}

For the seven model parameters ($\alpha_A$, $\alpha_S$, $\alpha_R$, $\alpha_C$, $\beta_2$, $\beta_3$, $\beta_4$, while $\beta_1$ is fixed to 1), we want to find a configuration that minimizes both mean $h$-index error and mean precision error:
\begin{equation}
R^{h}_{error} = \langle 1 - R^{h} \rangle, \\ 
P_{error} = \langle (1-P)\sqrt{|K|} \rangle \label{eq:GSP_rec_prec_error}.
\end{equation}
This mean $P_{error}$ can be artificially small because it is averaged over (mostly)  small clusters which easily achieve high precision. Hence, in the definition of our optimization scheme we introduce a counterbalancing statistical weight that accounts for size by requiring the algorithm to preferentially optimize the large clusters due to the cost incurred if any large cluster's precision error value, $1-P$,  is high. Relying on basic statistical arguments, the natural weight that we should give the large clusters is the statistical fluctuation scale attributable to size, which is proportional to square root of the size of the cluster. This weight also compensates for the fact that there are more smaller clusters than large clusters. In practice, this means that for two clusters of different sizes $K_{+}=f K_{-}$ (with $f>1$), then the larger cluster with $K_{+}$ will need to have a precision error equal to $(1-P_{-})/\sqrt f$ in order to contribute the same to the overall $P_{error}$ value which must be minimized by the algorithm.

Due to the simplicity of our algorithm, we can conduct an extensive sampling over the whole parameter space. The results in Fig.~\ref{fig:optimization} (a) show that there is a clear trade-off between the two types of errors and a lower limit that can be reached by our implementation. Our test data consists of 3,000 surnames that were randomly selected from WoS and where at least one profile could be found on Google Scholar. To further improve the result, we did an iterative local search on a 7-dimensional sphere around the best previous parameter configurations, starting with the best results from the random parameter sampling. For efficiency reasons and for cross-validation, we drew four random subsets with 500 surnames each and optimized them individually. In Fig.~\ref{fig:optimization} (a), we aim at an error that equally prefers a high $h$-index and precision correctness. We find
\begin{align*}
\alpha_A=0.54, \alpha_S=0.75, \alpha_R=0.19, \alpha_C=1.02, \beta_2=0.19, \beta_3=0.011, \beta_4=0.49
\end{align*}
which leads to a precision error of 11.84\% and an h-index error of 12.63\%. Co-authorship $\alpha_A$ comes out as a strong indicator for disambiguation, although co-author names are not disambiguated beforehand and hence represent a potential source of errors. Self-citations $\alpha_S$ are also highly weighted, but a self-citation link alone is not sufficient to exceed the threshold $\beta_1=1$ to form clusters. 

Fig.~\ref{fig:optimization} (b) shows how much the individual features (terms of Eq. (1)) contribute to the optimal solution. We fitted curves to the best results of a random sampling for a varying error trade-off, when only certain features are used (i.e. parameter of the other features are set to 0). Individual features cannot reach low error rates on their own. Combining features of the co-author and citation graph work best. Including more features like affiliations, topical features extracted from titles, summaries or keyword lists could potentially further improve the solution.

\begin{figure}
	\centering
	\subfigure[]{\label{fig:hindex}\includegraphics[width=0.475\textwidth]{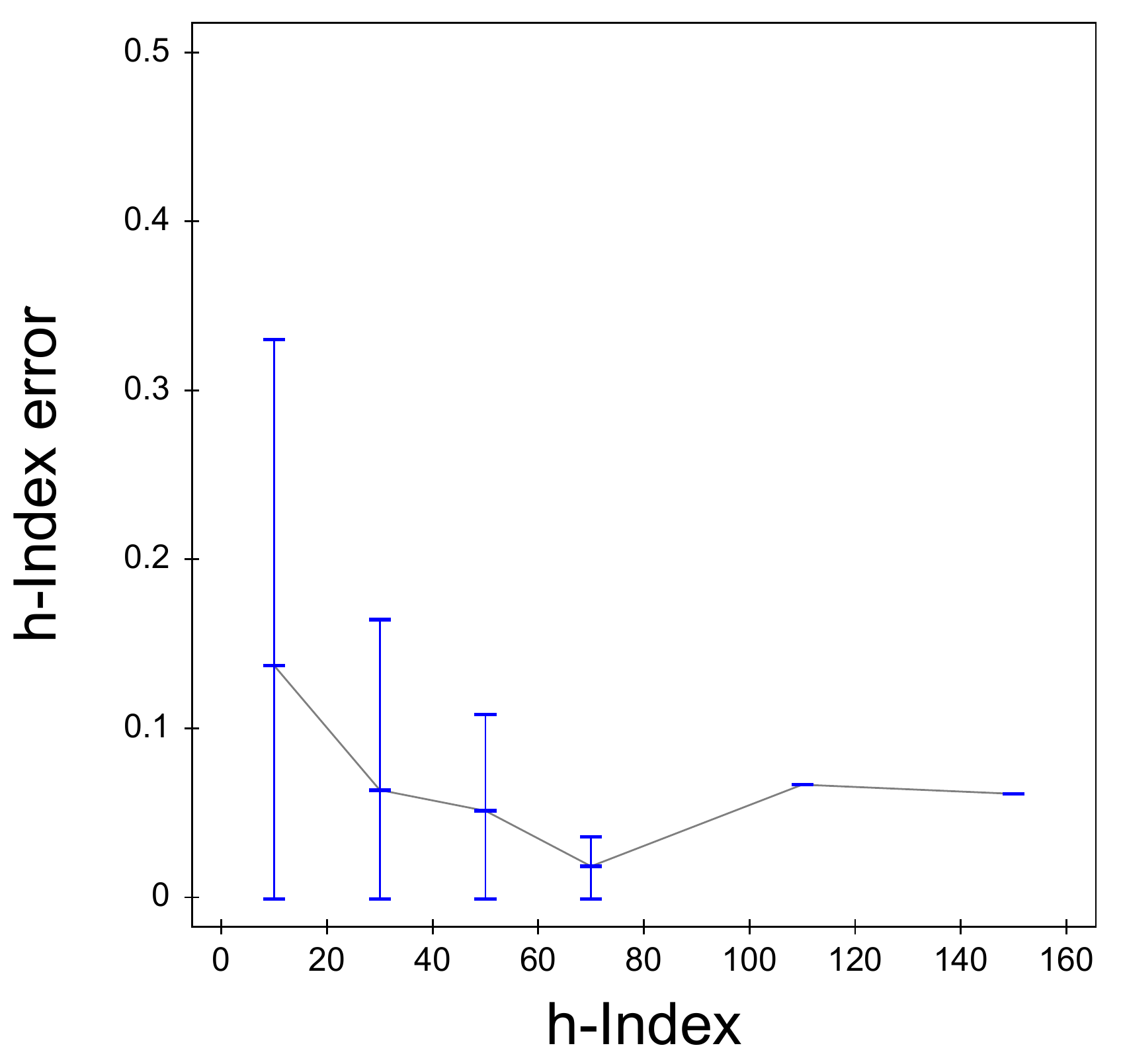}}\hfill
	\subfigure[]{\label{fig:precision}\includegraphics[width=0.475\textwidth]{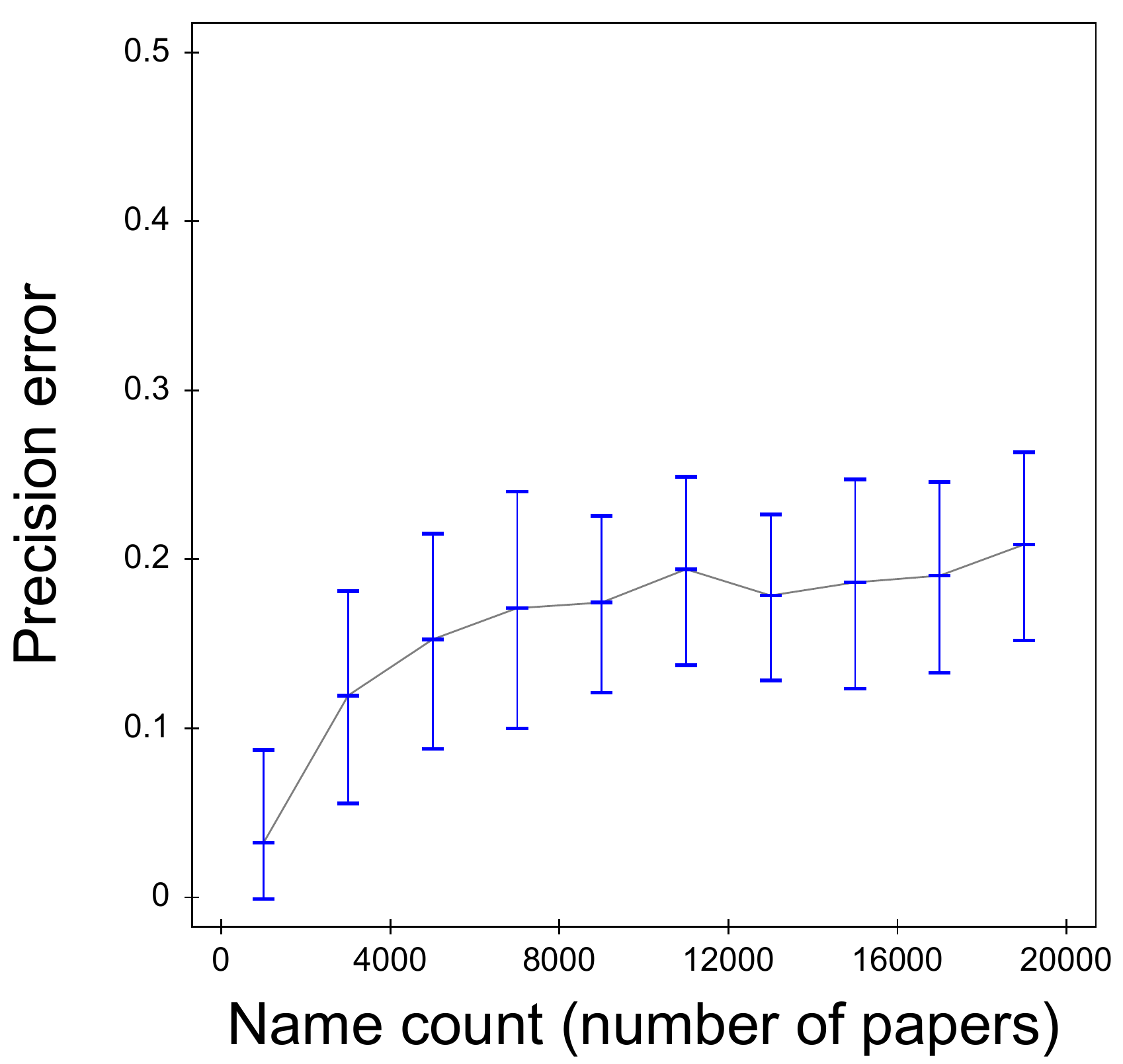}}
	\caption{ \textbf{Validation results of the 3,000 family names with an optimal parameter configuration.} \subref{fig:hindex} The mean $h$-index error (bin width=20, error bars displaying standard deviation) is descreasing for clusters with higher $h$-index. \subref{fig:precision} The precision error is increasing with more common names (bin width=2,000). }
	\label{fig:optimization2}
\end{figure}

Size dependent biases can skew aggregate algorithm performance measures especially when there is a broad underlying heterogeneity in the data. 
Hence, stating mean error rates is not sufficient to fully understand the strengths and weaknesses of a disambiguation algorithm. In Fig.~\ref{fig:optimization2} (a) we show that our algorithm works better for larger profiles, i.e. researchers that have a higher h-index, which is not a surprising result since there is much more co-author and citation graph information than for people with only a few papers. On the other hand, precision is slowly decreasing for more common names, see Fig.~\ref{fig:optimization2} (b), which becomes an issue when disambiguating very large databases, where certain combinations of surname plus first initial can result in initially undisambiguated clusters comprising around ten thousand publications.

\subsection{Further validation}

We further evaluated the performance of our disambiguation method with four additional tests using different data or techniques. While each measures recall or precision, these performance indicators have different definitions and deviate here from our previous validation, but fit better with measures typically reported in past disambiguation work.

We performed a manual disambiguation validation similar to the one in \citep{torvik2009author}. 100 publication pairs were randomly chosen from all pairs of publications that our algorithm co-clustered. Another 100 random pairs were selected from the set in which each pair belongs to the same name, but were placed in different clusters. Students were asked to determine for an author name and a given pair of publications, if they were written by the same author or different authors. When uncertain, the student could choose "Not sure". Although all resources could be used, this is often a challenging task and especially voting for "Different authors" frequently required evidence beyond that was easily available. From 138 answers, we obtained 111 "Same authors", of which 94 were in the same clusters (a recall of about 84.7\%), and 27 times "Different authors", of which all were correctly disambiguated to different clusters (a precision of 100\%). We point out that a manual disambiguation may be biased towards easy cases that could receive a confident answer, however, it does provide further evidence of the suitability of our algorithm.

Another test for precision can be constructed from second initials metadata which we do not consider for our disambiguation algorithm (only first initials when clustering the whole WoS). Indeed, about 4.7 million clusters contain at least two second initial names. Here, for each cluster the most common second initial forms the set of correctly disambiguated publications (names that omit the second initial were ignored). We measure a mean precision of about 95.4\%.

As a third way to evaluate precision, we "artificially" generated ground truth data by merging the sets of publications with two random names and then cluster them. The idea is that while we cannot say something about the correctness of the resulting clusters for one name, we can definitely show that the clustering is wrong when a cluster is generated from publications from both names. About 3,000 name pairs led to 26,887 clusters of which 18 clusters contained both names.

Our final additional validation is an estimate of recall, again for the whole disambiguated WoS. We evaluated about 870,000 arXiv.org publications, their metadata and fulltexts. From the PDFs more than half of all publications contained one or more email addresses. An email address is assumed to be a good indicator that, when two publications also share an author name, that this refers to the same unique researcher. Both arXiv and WoS provide DOIs for newer publications (starting around the year 2000), so cross-referencing was not an issue. We generated 110,011 "email" clusters, i.e. sets of publications that we also wanted to see for our disambiguation being put in one cluster. The mean recall was 98.1\%.

\subsection{Empirical h-index distribution and theoretical model}

\begin{figure}
	\centering
	\subfigure[]{\label{fig:hindexDistA}\includegraphics[width=0.49\textwidth]{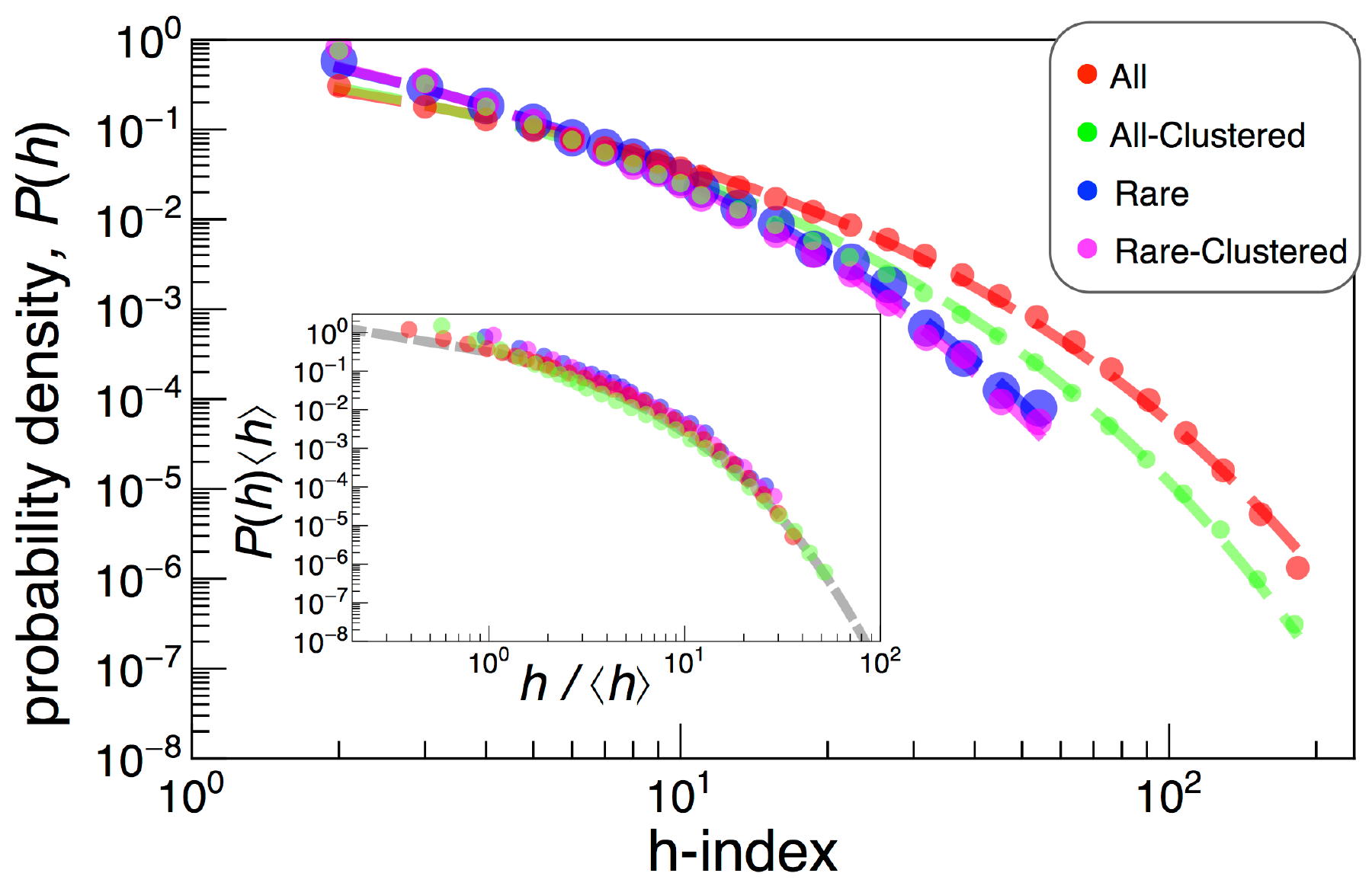}}\hfill
	\subfigure[]{\label{fig:hindexDistB}\includegraphics[width=0.46\textwidth]{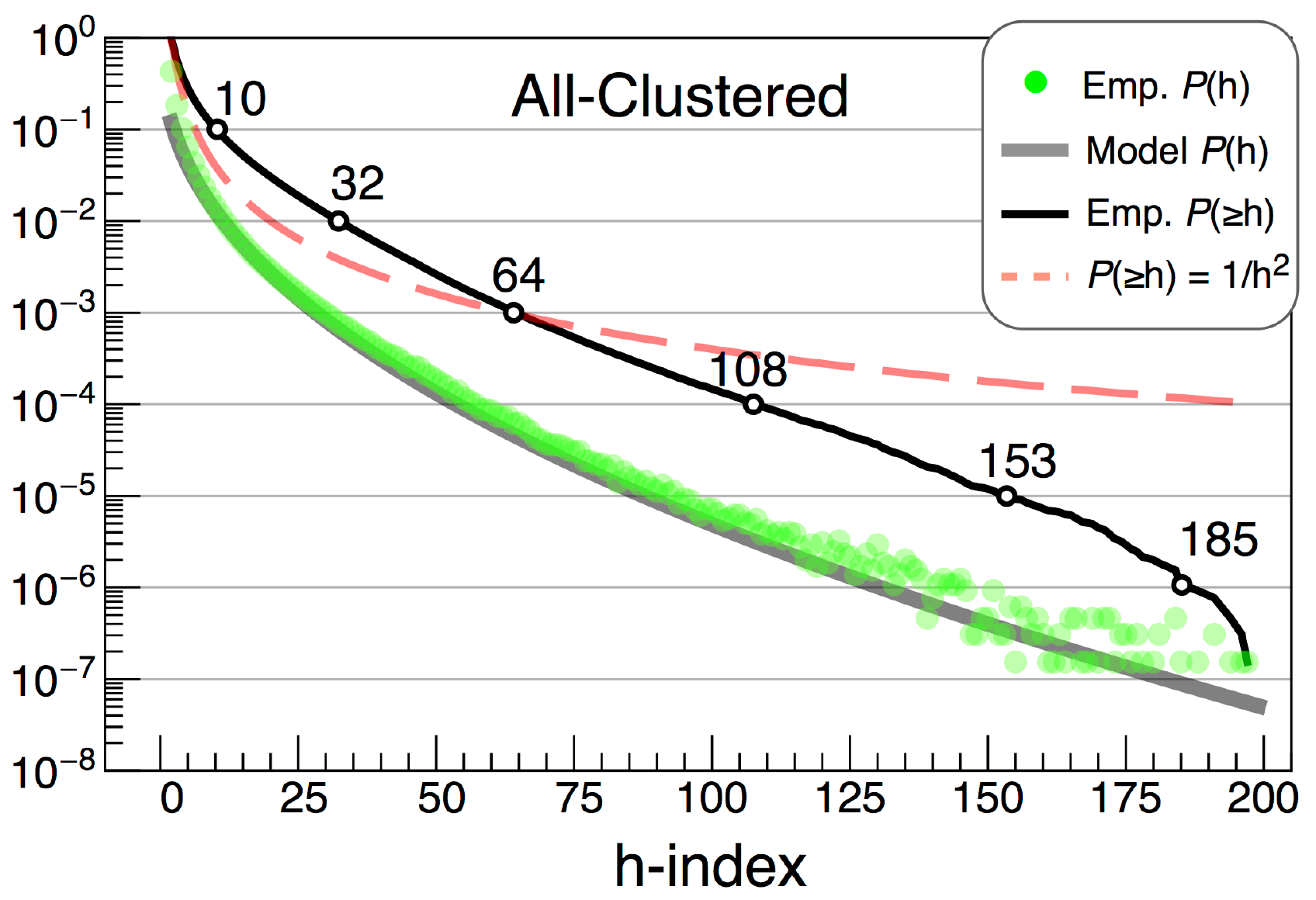}}
	\caption{ \textbf{Empirical and theoretical $h$-index distribution.} \subref{fig:hindexDistA} Testing the predictions of a stochastic  $h$-index model with empirical data. Shown for each dataset is the empirical probability density function $P(h)$, using logarithmic binning for $h>10$. We fit each $P(h)$ to the model distribution $P_{m}(h)$, parametrized by only the distribution average, which is  related to the mixing model parameters as $\langle h \rangle =\lambda_{1} \lambda_{2}$. (Inset) Data collapse of the empirical distributions along the universal curve $K_{0}(\sqrt{h}; \lambda_{1}\lambda_{2}=1)$ (dashed grey curve) using the scaled variable $x=h/\langle h \rangle$. \subref{fig:hindexDistB}  6,498,286 clusters with $h\geq2$ were identified for the entire WoS disambiguation. Plotted are the probability distribution $P(h)$ (green circles), the best-fit model $P_{m}(h)$ with $\lambda_{1}\lambda_{2}=3.49$, and the complementary cumulative distribution $P(\geq h)$ (solid black curve). The numbers indicate the value associated with the percentile $100\times(1-P(h))$, e.g. 1 per 1000 clusters (corresponding to the 99.9$th$ percentile) has $h$-index of 64 or greater.}
	\label{fig:hindexDist}
\end{figure}

Using the optimized parameters, we disambiguated the complete WoS database containing about 5.6 million author names that have a unique surname plus first initial.  
While the true $h$-index distribution is not exactly known, we can compare it to the subset of rare names - names for which we assume require little if any disambiguation. We define rare names as surnames where for the whole WoS there is only one type of initial and that initial is itself very rare (q, x, z, u, y, o, and w), which results in 87,000 author names. The disambiguation of the rare names tells us that they indeed represent to a large extent unique researchers. Unfortunately, for higher h-index values $h> 20$ (values in the top 3\% when excluding clusters with $h=0,1$) the rare surnames are underrepresented with respect to the whole database (see Fig. \ref{fig:hindexDist}  for the comparison between the rare dataset and the full dataset h-index distributions). However, this difference is consistent with deviations arising from finite-size effects, since the rare dataset is significantly smaller than the entire dataset.

The empirical distribution $P(h)$  is a mixture of  $h$-indices of scientists with varying discipline citation rates and varying longevity within mixed age-cohort groups. Hence, it may at first be difficult to interpret the mean value $\langle h \rangle$ as a representative measure for a typical scientist, since a typical scientist should be conditioned on career age and disciplinary factors. Nevertheless, in this  section we develop a simple mixing model that predicts the expected frequencies of $h$, hence providing insight into several  underlying features of the empirical ``productivity'' distribution $P(h)$.

Our $h$-index distribution model is based on the following basic assumptions:
\begin{enumerate}
\item The number of individuals of ``career age'' $t$ in the aggregate data sample is given by an exponential distribution $P_{1}(t) = \exp[-t/\lambda_{1} ]/\lambda_{1}$. We note that in this large-scale analysis we have not controlled for  censoring bias since a large number of the careers analyzed are not complete, and so the empirical data  likely overrepresent the number of careers with relatively small $t$.

\item The $h$-index growth factor $g_{i} \approx \langle h_{i}(t+1)- h_{i}(t) \rangle$ is the characteristic annual change in $h_{i}$ of a given scientist, and is distributed according to an exponential distribution $P_{2}(g) = \exp[-g / \lambda_{2}]/\lambda_{2}$. The quantity $g$ captures unaccounted factors such as the author-specific citation rate (due to research quality, reputation, and other various career factors), as well as the  variation in  citation and publication  rates across discipline. For sake of simplicity, we assume that $g_{i}$  is uncorrelated with $t_{i}$.
\end{enumerate}
Hence, the index $h_{i} = g_{i} t_{i}$ of an individual $i$ is simply given by the  product of a career age $t_{i}$ and growth factor $g_{i}$. 
The aggregate $h$-index distribution model $P_{m}(h)$ is derived from the distribution of a product of two random variables, $t$ and $g$, each distributed exponentially by $P_{1}(t;\lambda_{1})$ and $P_{2}(g;\lambda_{2})$, respectively. Since both $g\geq0$ and $t>0$, the distribution $P(h)$ is readily calculated by
\begin{eqnarray}
P_{m}(h) = \int_{0}^{\infty} \frac{dx}{x} P_{1}(x) P_{2}(h/x) \nonumber = \frac{2}{\lambda_{1}\lambda_{2}} K_{0}\Big(2\sqrt{h/(\lambda_{1}\lambda_{2})}\Big) \ , 
\label{compPDF2}
\end{eqnarray}
where $K_{0}(x)$ is the Modified Bessel function of the second kind. The probability density function $P_{m}(h)$ has mean $\langle h \rangle = \lambda_{1}\lambda_{2}$, standard deviation $\sqrt{3} \langle h \rangle$, and asymptotic behavior  $P_{m}(h) \sim \exp[-\sqrt{h/\langle h \rangle}]/h^{1/4}$ for $h\gg1$. 

Fig. \ref{fig:hindexDist}(A) shows the empirical distribution $P(h)$ for 4 datasets, analyzing only clusters with $h\geq 2$ in order to focus on clusters that have at least two cited papers which satisfy our similarity threshold with at least one other paper. Surprisingly, each $P(h)$ is well fit by the theoretical model $P_{m}(h;\lambda_{1}\lambda_{2})$ with varying $\lambda_{1}\lambda_{2}$ parameter. The  $\lambda_{1}\lambda_{2}$ parameter value was calculated for each binned $P(h)$ using a least-squares method, yielding    $\lambda_{1}\lambda_{2} =$ 2.09 (Rare), 1.90 (Rare-Clustered), 5.13 (All), and 3.49 (All-Clustered).  
The inset demonstrates data collapse for all four $P(h/\langle h \rangle)$ distributions following from the universal scaling form of  $K_{0}(x)$. 

How do these findings compare with general intuition? Our empirical finding significantly deviates from  the prediction which follows from combining Lotka's productivity law \cite{Lotka1926fds}, which states that the number  $n$ of publications follows a Pareto power-law distribution $P_{p}(n) \sim n^{-2}$, and the  recent observation that the $h$-index scales as $h\sim n^{1/2}$ \cite{Castellano2013}, which together imply that $P_{p}(h)\sim h^{-3}$ (corresponding to $P_{p}(\geq h) = h^{-2}$). 

Fig. \ref{fig:hindexDist}(B) compares the empirical complementary cumulative distribution $P(\geq h)$ for both empirical data (representing the 6,498,286 clusters with $h\geq2$ identified by applying the disambiguation algorithm to the entire WoS dataset) and for the theoretical Pareto distribution $P_{p}(\geq h)=1/h^{2}$. There is a crossover between the two $P(\geq h)$ curves around $h\approx 64$ (corresponding to the $99.9th$ percentile) which indicates that for $h>64$ we observe significantly fewer clusters with a given $h$ value than predicted by Lotka's productivity law. For example, the Lotka law predicts a  100-fold increase in the number of scientific profiles with  $h$ larger than the 1 per million frequency, $h\geq 185$. This discrepancy likely reflects the finite productivity lifecycle of scientific careers, which is not accounted for in models predicting scale-free Pareto distributions.
 
So how do these empirical results improve our understanding of how the h-index should be used? We show that the sampling bias encountered in small-scale studies \cite{Hirsch2005}, and even large-scale studies \cite{Castellano2013}, significantly discounts the frequency of careers with relatively small $h$. We observe a monotonically decreasing $P(h)$ with a heavy tail, e.g. only 10\% of the clusters with $h\geq2$ also have $h \geq10$. This means that the $h$-index is a noisy comparative metric when $h$ is small since a difference $\delta h \sim 1$ can cause an extremely large change in any ranking between scientists in a realistic academic ranking scenario. 
Furthermore, our model suggests that disentangling the net $h$-index from its time dependent and discipline dependent factors leads to a more fundamental question: controlling for age and disciplinary factors, what is the distribution of $g$? Does the  distribution of $g$ vary dramatically across  age and disciplinary cohorts? This could provide better insight into the interplay between impact, career length \cite{petersen2011quantitative} and the survival  probability of academics \cite{Kaminksi2012,Petersen2012}.

\section{Conclusion}
\label{disc}

The goal of this work was to disambiguate all author names in the WoS database. We found that existing methods relied on metadata that are not available or not complete in WoS, or were not specifically developed for an application to such a huge database. Second, we needed a test dataset which is not limited to certain research fields or geographical regions, and large enough to be representative for WoS. As previous work had shown that even under less demanding conditions perfect disambiguation is not achievable, we concentrated on the most influential work to correctly disambiguate papers that are most cited.

We achieved our goal by disambiguating author names based on the citation graph, which is the main feature of WoS. This approach exploits the fact that, on average, there is much more similarity between two publications written by the same author than between two random publications from different authors who happen to have the same name. We maximized the separation between these two classes, which can be seen as positive or wanted links and unwanted links in a publication network that connects papers written by the same unique researcher. Counting shared outgoing references and incoming citations are a much more fine-grained disambiguation criterion than for example journal or affiliation entries. Our disambiguation method does not assume any specific feature distribution, but is parameterized and trainable according to a suitable ''gold standard". It turns out that Google Scholar author profiles, one of the emerging collections of user editable publication lists, can reasonably serve as such a standard.

Our proposed method consists of three main components that could be altered or improved while still keeping the same validation framework: the error measure, the similarity measure and the clustering algorithm. The error measure we presented was specifically developed for reproducing $h$-indices; we believe other goals could be accomplished as well. The similarity measure could be easily extended by further metadata. Furthermore, our clustering algorithm, while intuitive and computationally efficient, could potentially be replaced by some more sophisticated community detection.

Comparing our results with previous work is difficult, as there is no common benchmark available. There are several studies that analyze small subsets of authors names, which is certainly useful to understand the mechanisms of the respectively proposed algorithms and sometimes unavoidable in lack of a massive test dataset. We realized, however, that this does not allow for generalization across disciplines, time, career age, and varying metadata availability. We also point out that there are differences in the error reporting, mainly in the way how the mean of errors is calculated. The vast majority of authors has only one or two publications, making it likely that the low error rates for precision and recall are underestimated. Some publications report error rates lower than 1-2\%. We do not claim such an excellent result, since even our gold standard (cross-referenced publications from Google Scholar profiles, and name initials from WoS) cannot be assumed to have error rates significantly better than that. We have shown instead that using author and citation graph information only, we can disambiguate huge databases in a computationally efficient way and at the same time being flexible regarding the objectives one like to optimize for.

%%%%%%%%%%%%%%%%%%%%%%%%%%%%%%%%%%%%%%%%%%%%%%
%%                                          %%
%% Backmatter begins here                   %%
%%                                          %%
%%%%%%%%%%%%%%%%%%%%%%%%%%%%%%%%%%%%%%%%%%%%%%

\begin{backmatter}

\section*{Competing interests}
  The authors declare that they have no competing interests.

\section*{Author's contributions}
    DH proposed the citation-network-based name disambiguation approach. CS performed the data analysis. The h-index distribution model was developed by AP. The manuscript was written by CS, OP and AP.

\section*{Acknowledgements}
The authors would like to thank Tobias Kuhn and Michael Maes for useful feedback.  This work was inspired by the FuturICT project, in particular its part on the Innovation Accelerator. OP acknowledges funding from the Social Sciences and Humanities Research Council of Canada. We acknowledge Thomson ISI as required by the terms of use of our WoS data license.
%%  After submission of the .TEX file,                     %%
%%  you will be prompted to submit your .BBL file.         %%

% if your bibliography is in bibtex format, use those commands:
\bibliographystyle{bmc-mathphys} % Style BST file
\bibliography{biblio}      % Bibliography file (usually '*.bib' )

% or include bibliography directly:
% \begin{thebibliography}
% \bibitem{b1}
% \end{thebibliography}

%%%%%%%%%%%%%%%%%%%%%%%%%%%%%%%%%%%
%%                               %%
%% Figures                       %%
%%                               %%
%% NB: this is for captions and  %%
%% Titles. All graphics must be  %%
%% submitted separately and NOT  %%
%% included in the Tex document  %%
%%                               %%
%%%%%%%%%%%%%%%%%%%%%%%%%%%%%%%%%%%

%%%%%%%%%%%%%%%%%%%%%%%%%%%%%%%%%%%
%%                               %%
%% Additional Files              %%
%%                               %%
%%%%%%%%%%%%%%%%%%%%%%%%%%%%%%%%%%%

%\section*{Additional Files}

\end{backmatter}
\end{document}